\documentclass[final,5p,times,twocolumn,sort&compress]{elsarticle}
\usepackage{graphicx}
\usepackage{amsmath,amssymb}
\usepackage[final]{hyperref}
\hypersetup{
	colorlinks=true,       
	linkcolor=blue,        
	citecolor=blue,        
	filecolor=magenta,     
	urlcolor=blue         
}

\journal{Physics Letters A}

\begin{document}

\begin{frontmatter}


\title{Violation of the Bell inequality in quantum critical random spin-$1/2$ chains}

\author[1]{Jo\~{a}o C. Getelina}
\address[1]{Instituto de F\'{i}sica de S\~{a}o Carlos, Universidade de S\~{a}o Paulo, CP 369, 13560-970, S\~{a}o Carlos,
SP, Brazil}
\author[2]{Thiago R. de Oliveira}
\address[2]{Instituto de F\'{i}sica, Universidade Federal Fluminense, 24210-346, Niter\'{o}i, RJ, Brazil}
\author[1]{Jos\'{e} A. Hoyos}
\begin{abstract}
We investigate the entanglement and nonlocality properties of two random XX spin-1/2 critical chains, in order to better
understand the role of breaking translational invariance to achieve nonlocal states in critical systems. We show that breaking
translational invariance is a necessary but not sufficient condition for nonlocality, as the random chains remain in a local
ground state up to a small degree of randomness. Furthermore, we demonstrate that the random dimer model does not have the same
nonlocality properties of the translationally invariant chain, even though they share the same universality class for a certain
range of randomness.
\end{abstract}


\end{frontmatter}

\section{Introduction 
\label{sec:intro}}

The use of quantum information tools in condensed matter systems has become widespread, mostly because of their usefulness for a
better understanding of the behavior of quantum critical ground states (for a review, see Ref.~\cite{Amico2008}).
Currently, entanglement and nonlocality measurements are under intensive scrutiny since they have shown to be able to signal
quantum phase transitions\footnote{This is understood as a consequence of entanglement and nonlocality (as well as discord)
inheriting the nonanalytic behavior at the critical point from the usual spin-spin correlation functions} (QPTs) in many-body
systems~\cite{Batle2010,Justino2012,Deng2012,Wang2002,Sun2013,Sun2014,Sun2014B,Sun2014C,Huang2013}. Even though these concepts
are frequently associated with each other, it has been shown that they are indeed distinct by the construction of of entangled
mixed states which do not violate Bell-like inequalities\footnote{When we refer to Bell inequalities or Bell-like inequalities
we have in mind the original Bell inequality and the CHSH inequality (see Sec.~\ref{sec:fid-bell}).}~\cite{Werner1989}.
In addition, finding nonlocal states in many-body systems is of major interest, bearing in mind the many interesting
applications of nonlocal states, such as to cryptography~\cite{ekert-prl91} and to the generation of random
numbers~\cite{Pironio2010}.

Although it was observed that nonlocality measures may point out QPTs, it is far from clear what is the relation between nonlocality and QPTs. 
For instance, a recent study~\cite{Oliveira2012} has shown that due to monogamy and translational invariance, any
mixed state of a spin pair of the critical XXZ spin-1/2 chain is a local state, i.e., any spin pair does not violate the Bell inequality (even
though they can be in an entangled mixed state). This conclusion led us to inquire whether, generically, a
critical state is always local.

Therefore, we consider here two different spin-$1/2$ chains with randomly generated coupling constants.
By introducing randomness, we are able to break translational invariance without driving the system out of criticality.
In one these random models, the critical state belongs to the so-called infinite-randomness universality
class~\cite{Fisher1994}. In this case, when the degree of inhomogeneities is very large, there are spin pairs in nearly
Bell-like (singlet) states~\cite{Hoyos2006} which become strong candidates to violate the Bell inequality. In the other model,
the corresponding universality class is of finite-disorder type. It was shown that the corresponding ground state has
many similarities with the one of the translationally invariant case, such as sharing the same set of critical exponets (i.e.,
belonging to the same universality class) below a certain degree of randomness~\cite{Hoyos2011,Getelina2015}. It is then much
less clear whether the Bell inequality is violated or not.

We have shown here that the Bell inequality is violated in both cases, if the degree of randomness is greater than a
certain amount (which we have determined). Moreover, for the case in the infinite-randomness universality class, the spin pairs
violating the Bell inequality can be widely separated, while for the finite-disorder case only nearest-neighbor spin pairs may
be in nonlocal states. The most striking result is that the second model exhibits nonlocality even when it belongs to the same
universality class of the translationally invariant case (which was shown to be local).

The remainder of this paper is structured as follows: in Sec.~\ref{sec:XX} we present our random models, emphasizing the
differences between them. In Sec.~\ref{sec:fid-bell} we define and describe how to obtain the entanglement and nonlocality
measurements. Sec.~\ref{sec:data} presents our numerical results, which are further discussed in Sec.~\ref{sec:conclusion} and
followed by perspectives of future studies and applications.

\section{The random uncorrelated and correlated XX spin-$1/2$ models 
\label{sec:XX}}

Here, we introduce the two studied models, which are special cases of the disordered XXZ spin-$1/2$
chain~\cite{Masuda2004,Epstein1987}. This model is described by the Hamiltonian
\begin{equation}
H=\sum^{L}_{i=1}J_{i} \left( S^{x}_{i}S^{x}_{i+1} + S^{y}_{i}S^{y}_{i+1}  + \Delta_{i}S^{z}_{i}S^{z}_{i+1} \right),\label{eq:hamil}
\end{equation}
where $S^{\alpha}_{i}$ are the usual spin-$1/2$ operators, $J_{i}>0$ are the coupling constants, $\Delta_{i}$ are the 
anisotropy parameters and $L$ is the chain size which we will assume to be even. In addition, we will consider periodic boundary conditions: $\mathbf{S}_{i+L}=\mathbf{S}_{i}$.

In the translationally symmetric case ($J_i=J$ and $\Delta_i=\Delta$) the system is critical for $-1\leq \Delta \leq 1$ and
it is described as an exotic Tomonoga-Luttinger spin liquid state~\cite{giamarchi-book}, which is a highly
entangled~\cite{Amico2008} but local state, i.e., any spin pair does not violate the Bell
inequality~\cite{Justino2012,Oliveira2012}.

\begin{figure}
\begin{center}
\includegraphics[clip,scale=0.25]{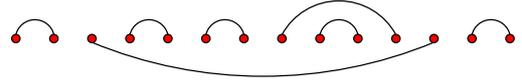}
\end{center}
\caption{\label{fig:rsp}Representation of the random singlet state where the dots are the spins in a regular lattice and
the curves connect spin pairs in nearly singlet states.}
\end{figure}

Conversely, in the uncorrelated random case ($J_i$ and $\Delta_i$ being uncorrelated and identically distributed random
variables) the system is described as a critical random singlet state for $-1/2<\Delta_i\leq 1$~\cite{Doty1992,Fisher1994} in
which spin pairs can be highly entangled in nearly singlet states~\cite{refael-moore-prl04,laflorencie-entanglement,Hoyos2006,Hoyos2007}, as depicted in Fig.~\ref{fig:rsp}. Remarkably, it was shown that this state is universal, in
the sense that all of its low-energy critical properties do not depend on (i) the details of the random variables, provided that
the width of their distribution is not zero and not unphysically large, and on (ii) the system anisotropy, provided that
$-1/2<\Delta_i\leq 1$.

For this reason, we here restrict our study to the case known as the XX model, in which $\Delta_i=0, \, \forall \, i$.
Another reason for our choice is due to the existence of a mapping between the XX chain and the tight-biding model of free
spinless fermions~\cite{Lieb1961}, which allows us to study considerably large chains via the exact diagonalization of the
Hamiltonian \eqref{eq:hamil}. 
Finally, it is plausible that our conclusions for the XX model also extend to the XXZ model in the
critical random-singlet region $-1/2<\Delta_{i}\le 1$ because, in this region, the ground state of the random XXZ chain depends
very weakly on the values of the local anisotropies $\Delta_i$, thus exhibiting the symmetry properties of the SU(2) symmetric
Heisenberg model $\Delta_i=1$~\cite{quito-hoyos-miranda-prl15}.

In our study, we draw the random couplings from a power-law like probability density distribution
\begin{equation}
P(J)=D^{-1}J^{\frac{1}{D}-1},\label{eq:dist}
\end{equation}
where $0<J<1$. Here, $D\geq0$ parameterizes the disorder strength, with $D=0$ recovering the translationally invariant case.
The probability distribution \eqref{eq:dist} is a natural choice as it allows us to assess a wide range of disorder strength
by varying the parameter $D$. Moreover, this probability distribution also coincides with the one of the infinite-randomness
fixed point, which governs the critical behavior of the system~\cite{Fisher1994}. Nonetheless, for the sake of completeness,
we have also considered the case of box-like distributions, i.e.,
\begin{equation}
P(J)= \begin{cases} 1, &\mbox{for } J_{\rm min} < J < 1 \\
0, & \mbox{otherwise} \end{cases}\label{eq:dist_box}
\end{equation}
In this case, $J_{\rm min}$ parameterizes the disorder strength, with smaller $J_{\rm min}$ meaning stronger disorder.

We now introduce our second model: the random correlated XX spin-$1/2$ chain.
The difference with respect to our first model is
that instead of considering an uncorrelated sequence of random couplings $\{J_1,\,J_2,\, \dots ,J_L\}$, we consider the special
sequence of couplings  $\{J_1,\,J_1,\,J_2,\,J_2,\, \dots ,J_{L/2},\,J_{L/2}\}$. Our interest in this special model is because it
was recently shown that short-range correlations among the random exchange couplings $J_i$ (e.g., the one we are considering
here, $J_{2i}=J_{2i-1}$) can dramatically change the low-energy properties of the XX spin-$1/2$
chain~\cite{binosi-entanglement-prb,Hoyos2011,Getelina2015}. 
For instance, the ground state of the random correlated model is
completely unrelated to the random-singlet state of the uncorrelated one; in fact, it even shares many similarities with the
ground state of the translationally invariant case. 
For $0\leq D\leq D_{\rm c}$, the ground-state bipartite (block) entanglement and the low-energy thermodynamics are
practically identical to those of the translationally invariant system~\cite{Getelina2015}.
Only for $D>D_{\rm c} \approx 0.3$ these quantities become distinct with, surprisingly, the (block) entanglement entropy
increasing with the disorder strength $D$ (and being greater than that of the translationally invariant)~\cite{Hoyos2011}.

\section{Entanglement and Violation of Bell Inequality 
\label{sec:fid-bell}}

In the strong-disorder limit ($D\gg1$), it is a good approximation to describe the ground state of \eqref{eq:hamil} (with uncorrelated random couplings) by the random-singlet state (see Fig.~\ref{fig:rsp}): a collection of independent singlets.
We now would like to test this approximation by measuring how far two spins $i$ and $j$ are from the actual singlet state 
$\left| \Psi^{-} \right\rangle=( \left|+-\right\rangle - \left|-+\right\rangle ) / \sqrt{2}$.  
For this reason, we study the so-called fidelity, which is given by 
\begin{equation}
F_{ij}=\left\langle \Psi^{-} \right| \rho_{ij} \left| \Psi^{-} \right\rangle,
\end{equation}
where $\rho_{ij}$ is the ground-state reduce density matrix encoding all the information about the physical state of the two spins $i$ and $j$.
Using the symmetries of the XX spin-$1/2$ chain Hamiltonian, one can related the fidelity to the ground-state transverse $C_{ij}^{xx}$ and longitudinal $C_{ij}^{zz}$ spin-spin correlation functions~\cite{Hoyos2006}: 
\begin{equation}
F_{ij}=\frac{1}{4}-2C^{xx}_{ij}-C^{zz}_{ij},\label{eq:fid}
\end{equation}
where $C^{\alpha\alpha}_{ij}=\left\langle S^{\alpha}_{i}S^{\alpha}_{j} \right\rangle = {\rm Tr}(\rho_{ij}S^{\alpha}_{i}S^{\alpha}_{j} )$. 
More importantly, the fidelity is related to the concurrence $\mathcal{C}_{ij}$ (a \textit{bona fide} entanglement measurement~\cite{Bennett1996,Hill1997,Wootters1998}) via
\begin{equation}
\mathcal{C}_{ij}= 
\begin{cases}
0, & \text{if $F_{ij} \le 1/2$}, \\
2F_{ij}-1, & \text{if $F_{ij} > 1/2$}.
\end{cases}
\end{equation}
Thus, for this model, the fidelity can be used as a entanglement measurement since it is monotonically related to the concurrence, with 
\begin{equation}
\label{eq:Fmin}
F_{ij}>1/2
\end{equation}
meaning that the two spins are entangled.

In addition to the entanglement, we also want to verify if the two-spins physical state is nonlocal by violating the Bell
inequality $\mathcal{B}_{ij}\leq 2$~\cite{Clauser1969,Horodecki1995}, where the Bell measurement for our model
Hamiltonian is simply~\cite{Justino2012} 
\begin{equation}
\mathcal{B}_{ij}=8\max \left\lbrace \sqrt{2 \left( C^{xx}_{ij} \right)^{2}}, \sqrt{\left( C^{xx}_{ij} \right)^{2}
+\left( C^{zz}_{ij} \right)^{2}} \right\rbrace.
\end{equation}
Moreover, for the XX spin-$1/2$ chain, we have verified that in all of our calculations $\left|C^{xx}_{ij}\right| \geq \left|C^{zz}_{ij}\right|$.\footnote{Although this assumption seems obvious, we were not able to prove it rigorously.
Further numerical inspections indicates that it is true for all cases in which $-1<\Delta_i<1$.}
Hence, the two-spin state is nonlocal whenever 
\begin{equation}
\label{eq:Cmin}
\left| C^{xx}_{ij} \right| > \frac{1}{4\sqrt{2}} \approx 0.1768.
\end{equation}
Notice furthermore that any spin pair which violates the Bell inequality is also entangled, but the reciprocal is not true.

As mentioned before, the nonviolation of the Bell inequality has been shown in the critical translationally invariant XXZ
spin-$1/2$ chain~\cite{Justino2012}. This property was later understood using the concept of monogamy in translationally invariant
systems~\cite{Oliveira2012} which, as a result, forbids that a given spin is in a nonlocal state with two neighbors
simultaneously. However, the introduction of disorder breaks translational invariance and, consequently, monogamy plays no
longer a role.

Finally, in order to gain further insight on the global structure of the entanglement properties of the ground state,
we study the monogamy relation $\sum_{j\neq i}\mathcal{C}_{ij}^{2} \le 1$ (for fixed $i$)~\cite{Coffman2000,Osborne2006,DeOliveira2014}.
Since we are dealing with random models, it is natural to sum over all sites $i$ and thus, we conveniently rewrite the monogamy relation as 
\begin{equation}
M=\frac{2}{L}\sum_{i=1}^{L-1}\sum_{j=i+1}^{L}\mathcal{C}_{ij}^{2} \le 1.\label{eq:mono}
\end{equation}
Closely related to this quantity is the total number of spin pairs violating the Bell inequality $Q_{\rm NL}$. 
As shown in Ref.~\cite{Oliveira2012}, three spins cannot be simultaneously in pairwise nonlocal states. Thus, $Q_{\rm NL}$
cannot be larger than the total number of spin pairs which is also $L/2$. 
For this reason, we evaluate the normalized quantity $2 Q_{\rm NL}/L \leq 1$

\section{Numerical results 
\label{sec:data}}

In this section, we present our numerical study of the entanglement and nonlocality properties of the two random models
introduced in Sec.~\ref{sec:XX}. In both cases, the random XX spin-$1/2$ chains are mapped into free spinless fermions from
which we can compute the ground-state transverse and longitudinal spin-spin correlation functions~\cite{Lieb1961}.
From these correlation functions, we are then able to compute the fidelity, concurrence, Bell measurement and monogamy, as
explained in Sec.~\ref{sec:fid-bell}. Except for the monogamy, due to the lack of translational invariance, these quantities do
not acquire a single distance-dependent value. Instead, they are randomly distributed and, for this reason, much more
information is gained from their probability density distributions. Therefore, we here compute the corresponding normalized
histograms.

We have considered spin chains of sizes $L=100$ and $200$ with periodic boundary conditions and verified that our data are
nearly free of finite-size effects. (In most cases, we have also checked this result by comparison with chains of size $L=400$).
For clarity, we will only show the results for $L=100$. The data were built considering $N=10^5$ different disorder
configurations for all the sizes studied, yielding an statistical error of a few percent.

\begin{figure}
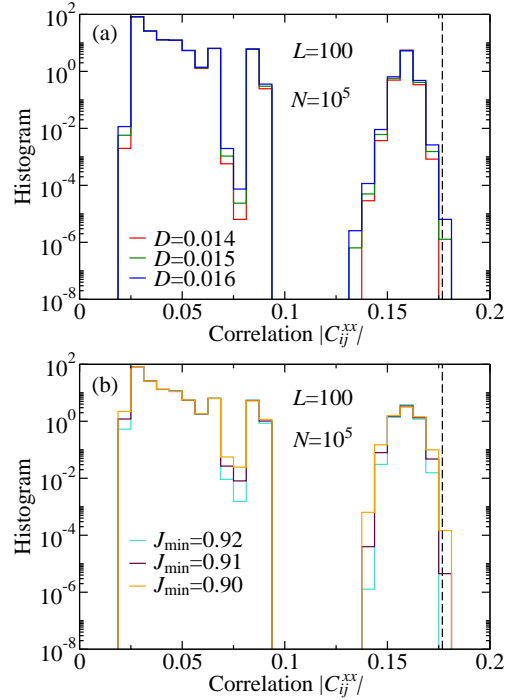

\begin{center}
\includegraphics[clip,scale=0.25]{fig2a}
\includegraphics[clip,scale=0.25]{fig2b}
\end{center}
\caption{\label{fig:uc-xc_v_small_ds}
Normalized histogram of the transverse correlation function $C^{xx}_{ij}$ for all possible spin pairs $i$ and $j$ for the
uncorrelated disorder model. In panel (a) the coupling constants are distributed in a power-law fashion [see
Eq.~\eqref{eq:dist}] while in panel (b) it is box-like distributed [see Eq.~\eqref{eq:dist_box}]. The dashed line represents
the nonlocality threshold
\eqref{eq:Cmin}.}
\end{figure}
\subsection{Uncorrelated disorder model}
We now present our results for the uncorrelated disorder spin-$1/2$ chain.

Our first goal is to investigate the minimum amount of disorder required for the existence of nonlocal states, i.e., spin
pairs which violate the Bell inequality. Fig.~\ref{fig:uc-xc_v_small_ds} shows the distribution of all spin pairs transverse
correlations $C_{ij}^{xx}$ for the cases in which the couplings are distributed (a) in a power-law \eqref{eq:dist} and (b) in a
box-like fashion. For disorder strength $D<D_{\rm NL}=0.015(1)$ (or $J_{\rm min}>J_{\rm NL}=0.91(1)$), with the number
in parenthesis denoting the statistical error of the last digit, there is no violation of the Bell inequality for any pair, and
thus, the state is local as in the translationally invariant case. Conversely, for $D>D_{\rm NL}$ (or
$J_{\rm min}<J_{\rm NL}$), we observe spin pairs that do violate the Bell inequality. We then arrive in one of our main
results, which is the existence of a nonlocal critical state in disordered systems.

\begin{figure}
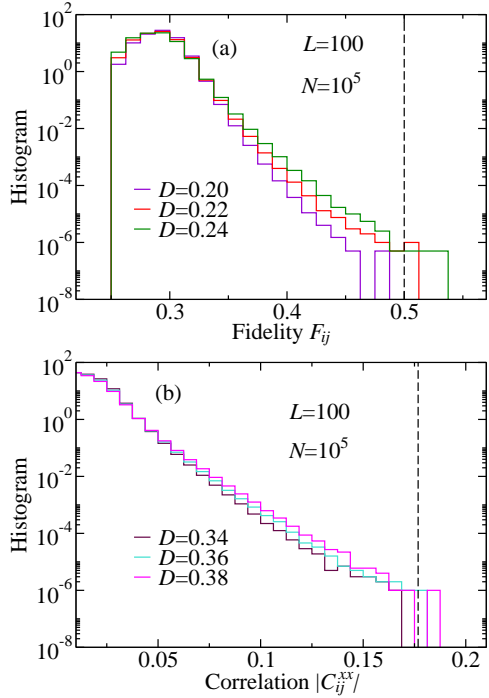

\begin{center}
\includegraphics[clip,scale=0.25]{fig3a}
\includegraphics[clip,scale=0.25]{fig3b}
\end{center}
\caption{\label{fig:uc-fid-lc_min_ds}
Normalized histogram of (a) fidelities $F_{ij}$ and (b) transverse correlation $C^{xx}_{ij}$ for the spin pairs $|i-j|>L/6$ for
the uncorrelated disorder model. The dashed lines represent the entanglement and nonlocality threshold, Eqs.~\eqref{eq:Fmin} and
\eqref{eq:Cmin}, respectively.}
\end{figure}

In addition, we would like to highlight one striking feature of the random singlet state (see Fig.~\ref{fig:rsp}): 
the spin pairs in a nonlocal state can be widely separated. In Fig.~\ref{fig:uc-fid-lc_min_ds} we show the distributions of
fidelities [panel (a)] and transverse correlation functions [panel (b)] for spin pairs widely separated from each other, namely,
$|i-j|>L/6$. We verify the existence of entangled pairs and nonlocality for $D>0.22(1)$ and $D>0.36(1)$, respectively.

\begin{figure}[t]
\begin{center}
\includegraphics[clip,scale=0.25]{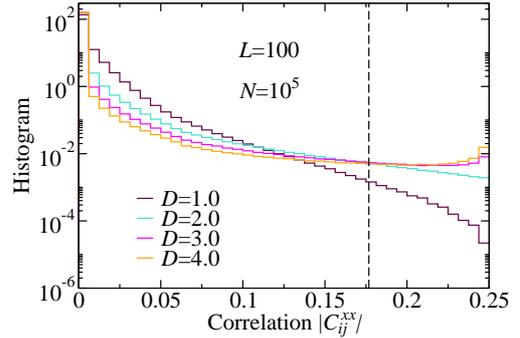}
\end{center}
\caption{\label{fig:uc-xc_v_ds}
Same as Fig.~\hyperref[fig:uc-fid-lc_min_ds]{\ref{fig:uc-fid-lc_min_ds}(b)} but for a different set of disorder parameter values $D$.}
\end{figure}

For completeness, we show in Fig.~\ref{fig:uc-xc_v_ds} the distribution of the transverse correlations for widely separated spin
pairs ($|i-j|>L/6$) for a broader range of disorder strength $D$. Clearly, the fraction of spin pairs violating the Bell
inequality increases with $D$. This behavior is expected since the random singlet state is known to become a better description
of the true ground state in the strong disorder regime~\cite{Hoyos2006}.

\begin{figure}[t]
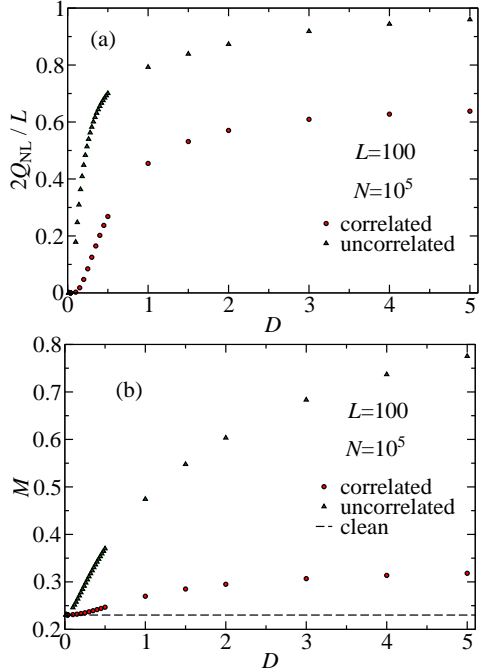

\begin{center}
\includegraphics[clip,scale=0.25]{fig5a}
\includegraphics[clip,scale=0.25]{fig5b}
\end{center}
\caption{\label{fig:nbel_v_ds} (a) The average number of spin pairs that violate the Bell inequality $Q_{\rm NL}$ and (b)
the monogamy relation for entanglement $M$ as a function of the disorder strength $D$ for both uncorrelated and correlated
disordered models. The dashed line is the value for the translationally invariant system (case $D=0$).}
\end{figure}

In order to further corroborate this result, we compute the average number of the spin pairs violating the Bell inequality
$Q_{\rm NL}$. In the random singlet state (as sketched in Fig.~\ref{fig:rsp}) there are $L/2$ singlet states, and thus, $L/2$
nonlocal spin pairs. Hence, it is expected that $2Q_{\rm NL}/L\rightarrow 1$ in the limit $D\rightarrow\infty$, as observed
in Fig.~\hyperref[fig:nbel_v_ds]{\ref{fig:nbel_v_ds}(a)}. Likewise, the monogamy relation \eqref{eq:mono} is expected to
saturate in the same limit, which is consistent with our results in Fig.~\hyperref[fig:nbel_v_ds]{\ref{fig:nbel_v_ds}(b)}.

\subsection{Correlated disorder case}
We now present our results for the case in which the couplings constants are correlated random variables appearing in pairs,
i.e., the sequence of random coupling constants is $\{J_1,\,J_1,\,J_2,\,J_2,\, \dots ,J_{L/2},\,J_{L/2}\}$, as explained in
Sec.~\ref{sec:XX}.

\begin{figure}
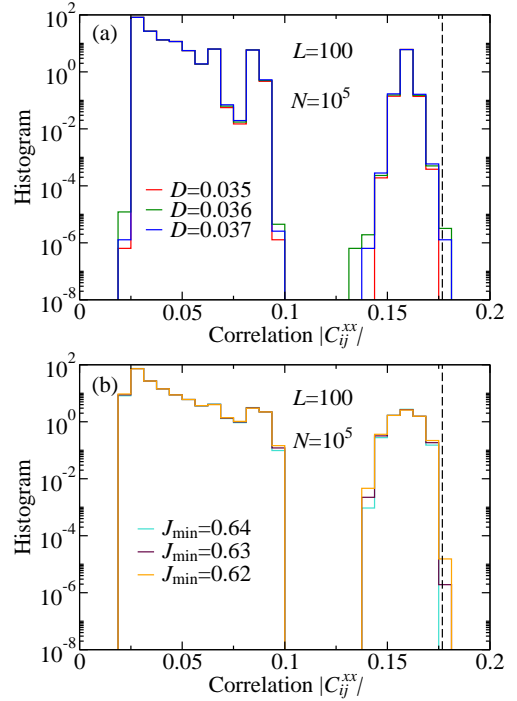

\begin{center}
\includegraphics[clip,scale=0.25]{fig6a}
\includegraphics[clip,scale=0.25]{fig6b}
\end{center}
\caption{\label{fig:cc-xc_v_small_ds}
Normalized histogram of the transverse correlation function $C^{xx}_{ij}$ for all possible spin pairs $i$ and $j$ for the
correlated disorder model. In panel (a) the coupling constants are distributed in a power-law fashion [see Eq.~\eqref{eq:dist}]
while in panel (b) it is box-like distributed [see Eq.~\eqref{eq:dist_box}]. The dashed line represents the nonlocality
threshold \eqref{eq:Cmin}.}
\end{figure}

Likewise the random uncorrelated model, we firstly investigate the minimum disorder strength necessary to have nonlocality.
As shown in Fig.~\ref{fig:cc-xc_v_small_ds}, nonlocality is obtained for $D>D_{\rm NL}^*=0.037(1)$ and
$J<J_{\rm NL}^*=0.62(1)$ for the cases of power-law [see Eq.~\eqref{eq:dist}] and box-like distributions, respectively.
This result is somewhat surprising, since as discussed before, the correlated random case is known to be in the same
universality class of the translationally invariant case for $D<D_{\rm c}\approx 0.3$~\cite{Hoyos2011,Getelina2015}. 
However, the translationally invariant case does not have spin pairs in a nonlocal state, which are, in contrast, already
observed for the correlated random case with $D_{\rm NL}^*<D<D_{\rm c}$. Thus, we have a critical model which exhibits
nonlocality even though it belongs to same universality class as a system with a local ground state.

\begin{figure}
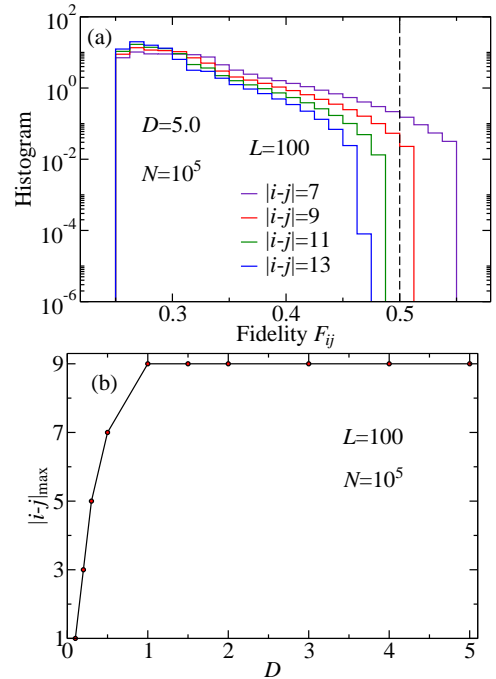

\begin{center}
\includegraphics[clip,scale=0.25]{fig7a}
\includegraphics[clip,scale=0.25]{fig7b}
\end{center}
\caption{\label{fig:cc-fid_r_max}
(a) Normalized histogram of fidelities $F_{ij}$ for the correlated disorder model, considering pairs with separations $|i-j|$
ranging from $7$ to $13$ and disorder strength $D=5.0$. The dashed line represents the entanglement threshold \eqref{eq:Fmin}.
(b) The maximum separation for entanglement $|i-j|_{\rm max}$ as a function of the disorder strength $D$.}
\end{figure}

We call the attention to the fact that the random correlated and uncorrelated models are fundamentally distinct. In order to see
this, we plot in Fig.~\hyperref[fig:cc-fid_r_max]{\ref{fig:cc-fid_r_max}(a)} the distribution of fidelities for spin pairs
separated by distances ranging from $7$ to $13$ in the high disordered case $D=5$. Differently from the random uncorrelated
case, only pairs with $|i-j| \leq 9$ are entangled. In Fig.~\hyperref[fig:cc-fid_r_max]{\ref{fig:cc-fid_r_max}(b)} we plot, as a
function of the disorder strength $D$, the maximum distance $|i-j|_{\rm max}$ for entangled spin pairs found among our
$N=10^5$ different chains for all system sizes studied. No entangled spin pair separated by distances greater than $9$ was found
for any disorder strength.

\begin{figure}
\begin{center}
\includegraphics[clip,scale=0.25]{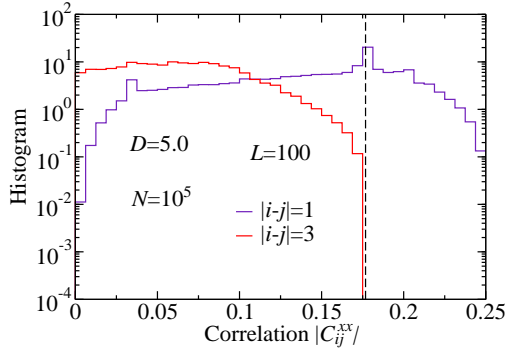}
\end{center}
\caption{\label{fig:cc-xc_r_max}
Normalized histogram of transverse correlation $C^{xx}_{ij}$ for the correlated disorder model, considering only pairs with
distance $|i-j|=1$ and $3$ and disorder strength $D=5.0$.  The dashed line represents the nonlocality threshold
\eqref{eq:Cmin}.}
\end{figure}

In addition, the spin pairs violating the Bell inequality are only the nearest-neighbor ones, as shown in
Fig.~\ref{fig:cc-xc_r_max}. With these results, there is no doubt about the fundamental difference between the ground states of the correlated and uncorrelated random models.

Finally, we show in Figs.~\hyperref[fig:nbel_v_ds]{\ref{fig:nbel_v_ds}(a)} and \hyperref[fig:nbel_v_ds]{(b)} the average number
of spin pairs violating the Bell inequality $Q_{\rm NL}$ and the monogamy $M$ [see Eq.~\eqref{eq:mono}] as a function of the
disorder strength $D$, respectively. Although these quantities increase with $D$, they do not saturate near their corresponding
thresholds as in the uncorrelated disorder case.

\section{Discussion and Conclusion 
\label{sec:conclusion}}
We have studied the effects of disorder (i.e., spatial inhomogeneities described as random coupling constants) on the pairwise
entanglement and nonlocality properties of the ground state of two critical XX spin-$1/2$ chains. These two chains differ on the
nature of the random coupling constants; in one case they are correlated while in the other they are not. We have shown that
both random models have spin pairs violating the Bell inequality, which is in contrast with the translationally invariant case,
where the interplay with monogamy forbids the violation of Bell inequality. Thus, we have shown that critical random systems can
be nonlocal and have determined the minimum amount of disorder necessary. Therefore, it becomes clear that breaking
translational invariance is a necessary but not sufficient condition for nonlocality.

We have evaluated the entanglement and nonlocality properties of both random chains for a relatively wide disorder range,
considering all possible spin pairs and also fixed distances. As already expected, we have noticed many differences between the
uncorrelated and correlated disorder cases. For instance, the former requires less disorder than the latter to have pairs
violating the Bell inequality (see Figs.~\ref{fig:uc-xc_v_small_ds} and~\ref{fig:cc-xc_v_small_ds}).
Moreover, for the uncorrelated case, the spin pairs violating the Bell inequality can be widely separated, while for the
correlated case we have shown that only nearest neighbors can be in a nonlocal state (see Fig.~\ref{fig:cc-xc_r_max}).

The existence of widely separated nonlocal states for the uncorrelated disorder case can be readily understood as a feature of
the random singlet state, which gives a good description of the system ground state in the limit of strong disorder. 
The random singlet state consists of a collection of $L/2$ arbitrarily spaced singlets (see Fig.~\ref{fig:rsp}), where $L$ is
the chain length. Thus, as disorder is increased, one could expect the average number of pairs violating the Bell inequality
$Q_{\rm NL}$ for the uncorrelated case approaching the maximum value, as it is indeed verified in
Fig.~\hyperref[fig:nbel_v_ds]{\ref{fig:nbel_v_ds}(a)}. Furthermore, the monogamy relation~\eqref{eq:mono} exhibits a similar
behavior with disorder, as shown in Fig.~\hyperref[fig:nbel_v_ds]{\ref{fig:nbel_v_ds}(b)}, but the saturation takes place only
for higher disorder strengths.

However, for the correlated disorder case the system ground state cannot be described by the random singlet state and, thus,
the nonlocal states are not widely separated. In fact, for this case we have determined the maximum distances for entangled and
nonlocal states, which are $|i-j|_{\rm max}=9$ and $|i-j|_{\rm max}=1$, respectively. Moreover, the number of pairs
violating Bell and the monogamy relation also increase with disorder, but with
a saturation value far below the upper bounds. In addition, we would like to call the attention to a rather surprising result:
For disorder strength $D_{\rm NL}^*<D<D_{\rm c} \approx 0.3$, the system has spin pairs in
nonlocal states even though it is in the same universality class of the translationally invariant case (i.e., they have the same
set of critical exponents), which is known to be local~\cite{Oliveira2012}. Thus, critical systems in the same universality
class can have different nonlocality properties, which shows that nonlocality is not a universal property.

Finally, nonlocal states were recently found in many-body system of cold-atom system~\cite{Schmied2016,Tura2014}.
We expect our work to provide a useful reference for future experiments on random spin-$1/2$ chains.
\section*{Acknowledgements}
This work was supported by the Brazilian funding agencies CAPES, CNPq, FAPESP, and INCT-IQ.

\bibliographystyle{model1a-num-names}
\bibliography{references}

\end{document}